\documentclass[twocolumn, nofootinbib, aps, prd]{revtex4}

\usepackage[dvipdfm, colorlinks=true, citecolor=blue, linkcolor=blue, urlcolor=blue]{hyperref}
\usepackage{graphicx}
\usepackage{dcolumn}
\usepackage{bm}
\usepackage{hyperref}
\usepackage{color}
\usepackage{amsfonts}
\usepackage{graphicx,amsfonts,multirow}
\usepackage{amsmath}
\newcommand{\DS}[1]{/\!\!\!#1}

\allowdisplaybreaks

\begin{document}

\title{The $D_s$-meson leading-twist distribution amplitude within the QCD sum rules and its application to the $B_s \to D_s$ transition form factor}

\author{Yi Zhang$^{1}$}
\email{yizhangphy@cqu.edu.cn}
\author{Tao Zhong$^{2}$}
\email{zhongtao1219@sina.com}
\author{Hai-Bing Fu$^{2}$}
\email{fuhb@cqu.edu.cn}
\author{Wei Cheng$^{3}$}
\email{chengw@cqu.edu.cn}
\author{Xing-Gang Wu$^{1}$}
\email{wuxg@cqu.edu.cn}

\address{$^1$ Department of Physics, Chongqing University, Chongqing 401331, People's Republic of China\\
$^2$ Department of Physics, Guizhou Minzu University, Guiyang 550025, People's Republic of China\\
$^3$ State Key Laboratory of Theoretical Physics, Institute of Theoretical Physics, Chinese Academy of Sciences, Beijing, 100190, P.R. China}

\date{\today}

\begin{abstract}

We make a detailed study on the $D_s$ meson leading-twist LCDA $\phi_{2;D_s}$ by using the QCD sum rules within the framework of the background field theory. To improve the precision, its moments $\langle \xi^n\rangle _{2;D_s}$ are calculated up to dimension-six condensates. At the scale $\mu = 2{\rm GeV}$, we obtain: $\langle \xi^1\rangle _{2;D_s}= -0.261^{+0.020}_{-0.020}$, $\langle \xi^2\rangle _{2;D_s} = 0.184^{+0.012}_{-0.012}$, $\langle \xi^3\rangle _{2;D_s} = -0.111 ^{+0.007}_{-0.012}$ and $\langle \xi^4\rangle _{2;D_s} = 0.075^{+0.005}_{-0.005}$. Using those moments, the $\phi_{2;D_s}$ is then constructed by using the light-cone harmonic oscillator model. As an application, we calculate the transition form factor $f^{B_s\to D_s}_+(q^2)$ within the light-cone sum rules (LCSR) approach by using a right-handed chiral current, in which the terms involving $\phi_{2;D_s}$ dominates the LCSR. It is noted that the extrapolated $f^{B_s\to D_s}_+(q^2)$ agrees with the Lattice QCD prediction. After extrapolating the transition form factor to the physically allowable $q^2$-region, we calculate the branching ratio and the CKM matrix element, which give $\mathcal{B}(\bar B_s^0 \to D_s^+ \ell\nu_\ell) = (2.03^{+0.35}_{-0.49}) \times 10^{-2}$ and $|V_{cb}|=(40.00_{-4.08}^{+4.93})\times 10^{-3}$.

\end{abstract}

\maketitle

\section{introduction}

Since the first measurement of the ratio $\mathcal{R}(D^{(\ast)})$ of the branching fractions $\mathcal{B}(B\to D^{(\ast)}\tau\nu_\tau)$ and $\mathcal{B}(B\to D^{(\ast)}\ell \nu_{\ell})$, where $\ell$ stands for the light lepton $e$ or $\mu$, had been reported by the BaBar Collaboration, the $B\to D^{(\ast)}$ semileptonic decays have attracted great attentions due to large differences between the experimental measurements~\cite{BABAR_Lees:2012xj, BABAR_Lees:2013uzd, BELLE_Huschle:2015rga, HFAG_Amhis:2014hma} and the standard model (SM) predictions~\cite{LCSR_Zuo:2006dk, Zuo:2006re, HQET_Fajfer:2012vx, Fu:2013wqa, PQCD_Fan:2013qz, PQCD_Fan:2015kna, LQCD_Lattice:2015rga, LQCD_Na:2015kha, Zhong:2018exo, Wang:2017jow}. Such difference has been considered as an evidence of new physics. Comparing with the $B^{0, +}$ decays, because its background contamination from the partial reconstruction decay could be less serious, the $B_s\to D_s \ell\nu_\ell$ decay is experimentally attractive. A natural question is whether there is also evidence of new physics in the semileptonic decay $B_s\to D_s \ell\nu_\ell$. This decay could also be an important channel for determining the Cabibbo-Kobayashi-Maskawa (CKM) matrix element $|V_{\rm cb}|$.

The LHCb collaboration reported the measurement of $|V_{\rm cb}|$ by using $B_s^0\to D_s^-\mu^+\nu_{\mu}$ and $B_s^0\to D_s^{\ast -}\mu^+\nu_{\mu}$ decays~\cite{Aaij:2020hsi}, in which the data of the proton-proton collision at the center-of-mass energies of $7$ and $8$ TeV with the integrated luminosity about $3~{\rm fb}^{-1}$ had been used in the analysis. By using the Caprini-Lellouch-Neubert (CLN) and the Boyd-Grinstein-Lebed (BGL) parameterization~\cite{Caprini:1997mu, Boyd:1994tt, Boyd:1995sq, Boyd:1997kz} for $B_s\to D_s$ transition form factor (TFF), the determined $|V_{\rm cb}|$ are $(41.4\pm 0.6\pm 0.9\pm 1.2)\times 10^{-3}$ and $(42.3\pm 0.8\pm 0.9\pm 1.2)\times 10^{-3}$, respectively. The LHCb collaboration also measured the ratio of the branching fractions $\mathcal{B}(B_s^0\to D_s^- \mu^+\nu_\mu)$ and $\mathcal{B}(B^0\to D^- \mu^+\nu_\mu)$, i.e. $\mathcal{R} = 1.09 \pm 0.05 \pm 0.06 \pm 0.05$, which then gives $\mathcal{B}(B_s^0\to D_s^- \mu^+\nu_\mu)=(2.49 \pm 0.12 \pm 0.14 \pm 0.16) \times 10^{-2}$.

The accuracy of theoretical predictions on the branching fraction $\mathcal{B}(B_s\to D_s \ell\nu_\ell)$ depends heavily on the TFF $f^{B_s\to D_s}_+(q^2)$. It has been calculated within several approaches, such as the quark models~\cite{Faustov:2012mt, Kramer:1992xr, Chen:2011ut}, the QCD light cone sum rules (LCSR)~\cite{Blasi:1993fi, Li:2009wq}, and the lattice QCD (LQCD)~\cite{Atoui:2013zza, Monahan:2017uby, McLean:2019qcx}. Similar to the $B\to\pi$ TFFs~\cite{Huang:2004hw}, the LQCD prediction is reliable in large $q^2$-region, the QCD factorization prediction or the quark model prediction is reliable in large recoil region $q^2\sim 0$, and the LCSR is reliable in low and intermediate $q^2$-regions. Predictions under various methods are complementary to each other. Because the LCSR prediction is applicable in a wider region and could be adapted for all $q^2$-region via proper extrapolations, and in this paper, we will adopt the LCSR approach to calculate $f^{B_s\to D_s}_+(q^2)$.

Generally, contributions from the light-cone distribution amplitude (LCDA) suffers from the power counting rules basing on the twists, i.e. the high-twist LCDAs are usually powered suppressed to the lower twist ones in large $Q^2$-region. The high-twist LCDAs may have sizable contributions to the LCSR, and how to ``design" a proper correlator is a tricky problem for the LCSR approach. By choosing a proper correlator, one can not only study the properties of the hadrons but also simplify the theoretical uncertainties effectively. As the usual treatment, the correlator is constructed by using the currents with definite quantum numbers, such as those with definite $J^P$, where $J$ is the total angular momentum and $P$ is the parity of the bound state. Such a construction of the correlator is not the only choice suggested in the literature, e.g. the chiral correlator with a chiral current in between the matrix element has also been suggested to suppress the hazy contributions from the uncertain LCDAs~\cite{Huang:1998gp, Huang:2001xb, Wan:2002hz, Zuo:2006dk, Wu:2007vi, Wu:2009kq}. In the paper, we adopt a chiral correlator to do the LCSR calculation, and we shall find that the leading-twist LCDA $\phi_{2;D_s}$ provides dominant contributions. Therefore, if an accurate $\phi_{2;D_s}$ has been achieved, we shall obtain an accurate prediction on $f^{B_s\to D_s}_+(q^2)$.

Till now, there are few calculations on the $D_s$-meson leading-twist LCDA $\phi_{2;D_s}$; recently, it has been studied by using the light-front quark model~\cite{Dhiman:2019ddr}. We shall first construct a light-cone harmonic oscillator model for $\phi_{2;D_s}$ based on the well-known BHL-description~\cite{BHL1, BHL2, BHL3} as we have done for $\pi$, $\rho$, $D$ and heavy meson LCDAs~\cite{BHL_Zhong:2014jla, BHL_Zhong:2014fma, BHL_Zhong:2016kuv, BHL_Zhang:2017rwz, BHL_Zhong:2018exo, Fu:2014cna, Fu:2014pba, Zeng:2021hwt}. Then its input parameters shall be fixed by using reasonable constraints such as the probability of finding the leading Fock-state in $D_s$-meson Fock-state expansion, the normalization condition, and the calculated LCDA moments $\langle \xi^n\rangle _{2;D_s}$ or the Gegenbauer moments $a_n^{D_s}$. All those moments shall be computed by using the QCD sum rules~\cite{SVZ_Shifman:1978bx} within the framework of background field theory (BFT)~\cite{BFT_Huang:1989gv} up to dimension-six operators.

The remaining parts are organized as follows. The LCSR for $B_s \to D_s$ TFF, the QCD sum rules of the moments of $\phi_{2;D_s}$ and the light-cone harmonic oscillator model for $\phi_{2;D_s}$ are given in Sec.\ref{Sec:II}. Numerical results and discussions are presented in Sec.\ref{Sec:III}. Section \ref{Sec:IV} is reserved for a summary. The useful functions for calculating the $\phi_{2;D_s}$ moments are listed in the Appendix.

\section{Calculation Technology}\label{Sec:II}

\subsection{The LCSR for $B_s \to D_s$ TFF}\label{Calc:TFF}

The $B_s\to D_s$ TFF $f^{B_s\to D_s}_+(q^2)$ and $\tilde{f}^{B_s\to D_s}(q^2)$ are usually defined as:
\begin{eqnarray}
&& \langle D_s(p)| \bar{c}\gamma_\mu b |B_s(p+q)\rangle
\nonumber\\
&& \quad\quad = 2p_\mu f^{B_s\to D_s}_+(q^2)  + q_\mu \tilde{f}^{B_s\to D_s}(q^2) ,
\label{TFFDefinition}
\end{eqnarray}
where $p$ is the momentum of $D_s$-meson, $q$ is the momentum transfer. In this paper, we focus on the semileptonic decay $B_s\to D_s \ell \bar{\nu}_{\ell}$ with $\ell = (e,\mu)$. The masses of light-leptons are negligible, and then due to chiral suppression, only $f^{B_s\to D_s}_+(q^2)$ is relevant for our present analysis.

To derive the LCSR of $f^{B_s\to D_s}_+(q^2)$, we adopt the following chiral correlation function (correlator):
\begin{widetext}
\begin{eqnarray}
\Pi_\mu(p,q) &=& i\int d^4x e^{ip\cdot x} \langle D_s(p)| T \left\{ \bar{c}(x)\gamma_\mu (1+\gamma_5)b(x), \bar{b}(0) i(1+\gamma_5)s(0) \right\} |0\rangle  \nonumber\\
&=& \Pi\left[ q^2, (p+q)^2 \right] p_\mu + \tilde{\Pi} \left[ q^2, (p+q)^2 \right] q_\mu.
\label{TFFCorrelator}
\end{eqnarray}
\end{widetext}

The correlator is analytic in whose $q^2$-region. In the timelike region, by inserting a complete series of the intermediate hadronic states into the correlator, one can obtain its hadronic representation by isolating out the pole term of the lowest stat of the $B_s$ meson. By further using the TFF definition (\ref{TFFDefinition}) and the $B_s$ meson decay constant $f_{B_s}$
\begin{equation}
\langle B_s|\bar{b}i\gamma_5 s|0\rangle  = m_{B_s}^2 f_{B_s}/m_b,
\end{equation}
where $m_{B_s}$ is the $B_s$-meson mass and $m_b$ is the $b$-quark mass, then the hadronic representation for the correlator (\ref{TFFCorrelator}) reads
\begin{eqnarray}
\Pi^{\rm had}\left[ q^2,(p+q)^2 \right] &=& \frac{2f^{B_s\to D_s}_+(q^2)m_{B_s}^2 f_{B_s}}{m_b\left[  m_{B_s}^2 - (p+q)^2 \right]} \nonumber\\
&+& \int^\infty_{s_0^{B_s}} \frac{\rho^{\rm QCD}(s)}{s-(p+q)^2} ds,
\label{HadronicRepresentation}
\end{eqnarray}
where $s^{B_s}_0$ is threshold parameter, $\rho^{\rm QCD}(s)$ is the spectral density, and we have implicitly used the conventional quark-hadron duality ansatz. On the other hand, in the spacelike region, the correlator can be calculated by using the operator production expansion (OPE) approach. It is done by using the $b$-quark propagator
\begin{eqnarray}
\langle 0|T b(x)\bar{b}(0)|0\rangle  = \int \frac{d^4k}{(2\pi)^4} e^{-ik\cdot x} \frac{\not\! k + m_b}{k^2 - m_b^2} + \cdots.
\label{bPropagator}
\end{eqnarray}
By matching the hadronic representation (\ref{HadronicRepresentation}) and the OPE of the correlator (\ref{TFFCorrelator}) with the help of the dispersion relation, the LCSR of $f^{B_s\to D_s}_+(q^2)$ can be obtained
\begin{eqnarray}
f_+^{B_s\to D_s}(q^2) &=& \frac{e^{m_{B_s}^2/M^2}}{m_{B_s}^2 f_{B_s}} \left[ F_0(q^2,M^2,s_0^{B_s}) \right. \nonumber\\
&+& \left. \frac{\alpha_s C_F}{4\pi} F_1(q^2,M^2,s_0^{B_s}) \right],
\label{f+_LCSR}
\end{eqnarray}
where $C_F = 4/3$, $M$ is the Borel parameter. Here the Borel transformation has been adopted to suppress continuum contributions. The leading-order (LO) contribution of $f_+^{B_s\to D_s}(q^2)$ takes the form:
\begin{eqnarray}
F_0(q^2,M^2,s_0^{B_s}) &=& \frac{m_b^2 f_{D_s}}{m_{B_s}^2 f_{B_s}} e^{m_{B_s}^2/M^2} \int^1_\Delta \frac{du}{u} \phi_{2;D_s}(u) \nonumber\\
&& \times\exp \left[ -\frac{m_b^2 - \bar{u} (q^2 - u m_{D_s}^2)}{u M^2} \right],
\label{LO}
\end{eqnarray}
with the $D_s$-meson decay constant $f_{D_s}$ and
\begin{eqnarray}
\Delta &=& \frac{1}{2m_{D_s}^2} \left[ \sqrt{(s_0^{B_s} - q^2 - m_{D_s}^2)^2 + 4m_{D_s}^2 (m_b^2 - q^2)} \right. \nonumber\\
&& \left. -(s_0^{B_s} - q^2 - m_{D_s}^2) \right].  \nonumber
\end{eqnarray}
The NLO contribution of $f_+^{B_s\to D_s}(q^2)$ reads
\begin{align}
&
F_1(q^2,M^2,s_0^{B_s})
\nonumber\\
&
= \frac{f_{D_s}}{\pi} \int^{s_0^{B_s}}_{m_b^2} ds e^{-s/M^2} \int^1_0 du {\rm Im}_s T_1(q^2, s, u) \phi_{2;D_s}(u). \nonumber\\
\label{NLO}
\end{align}
The imaginary part of the next-to-leading order amplitude $T_1$ can be read from Ref.\cite{Duplancic:2008ix}. Due to the present choice of the chiral correlator (\ref{TFFCorrelator}), contributions from the twist-3 $D_s$-meson LCDA exactly vanish in the LCSR. Thus the terms from omitted gluonic field in $b$-quark propagator (\ref{bPropagator}) and hence contributions from even higher-twist terms are negligibly small and can be safely neglected. Our remaining task is then to achieve a precise $\phi_{2;D_s}$.

\subsection{Sum rules for the moments of the $D_s$-meson leading-twist LCDA $\phi_{2;D_s}$}

The $D_s$-meson leading-twist LCDA $\phi_{2;D_s}$ is defined as
\begin{align}
&
\langle 0|\bar{c}(z)\DS z\gamma_5 s(-z) |D_s(q)\rangle
\nonumber\\
& \qquad\quad
= i(z\cdot q) f_{D_s} \int^1_0 dx e^{i(2x-1)(z\cdot q)} \phi_{2;D_s}(x),
\label{DADefinition}
\end{align}
where $f_{D_s}$ is the $D_s$ meson decay constant. The moments of $\phi_{2;D_s}(x)$ can be derived by expanding the left-hand-side of Eq.(\ref{DADefinition}) around $z=0$ and the exponent in the right-hand-side of Eq.(\ref{DADefinition}) as a power series, e.g.
\begin{align}
\langle 0|\bar{c}(0)\DS z\gamma_5 (z\cdot \tensor{D})^{n} s(0) |D_s(q)\rangle
= i f_{D_s}(z\cdot q)^{n+1} \langle \xi^{n} \rangle_{2;D_s},
\end{align}
where the $n_{\rm th}$-moment is defined as
\begin{eqnarray}
\langle \xi^n\rangle _{2;D_s} = \int^1_0 dx (2x-1)^n \phi_{2;D_s}(x).
\label{MomentDefinition}
\end{eqnarray}
The $0^{\rm th}$-moment satisfies the normalization condition
\begin{equation}
\langle \xi^0\rangle _{2;D_s} = 1.
\label{0Moment}
\end{equation}

The sum rules of those moments can be derived by using the following correlator
\begin{eqnarray}
\Pi^{(n,0)}_{2;D_s} (z,q) &=& i \int d^4x e^{iq\cdot x} \langle 0| T \{ J_n(x) J^\dag_0(0) \} |0\rangle
\nonumber\\
&=& (z\cdot q)^{n+2} I^{(n,0)}_{2;D_s} (q^2) ,
\label{correlator}
\end{eqnarray}
where $n=(0,1,2,...)$, and the currents
\begin{eqnarray}
J_n(x) &=& \bar{c}(x) {z\!\!\!\slash} \gamma_5 (i z\cdot \tensor{D})^n s(x), \\
J^\dagger_0(0) &=& \bar{s}(0) {z\!\!\!\slash} \gamma_5 c(0).
\end{eqnarray}
By applying the OPE for the correlator (\ref{correlator}) in deep Euclidean region based on the BFT~\cite{BFT_Huang:1989gv}, we obtain
\begin{align}
&\Pi^{(n,0)}_{2;D_s} (z,q) = i \int d^4x e^{iq\cdot x}
\nonumber\\
&\qquad\times  \left\{ - {\rm Tr} \langle 0| S_F^c(0,x) {z\!\!\!\slash} \gamma_5 (i z\cdot \tensor{D})^n S_F^s(x,0) {z\!\!\!\slash} \gamma_5  |0\rangle  \right.
\nonumber\\
&\qquad
+ \left. {\rm Tr} \langle 0| S_F^c(0,x) {z\!\!\!\slash} \gamma_5 (i z\cdot \tensor{D})^n \bar{s}(0)s(x) {z\!\!\!\slash} \gamma_5  |0\rangle  \right\} \nonumber\\
&\qquad
+ \cdots,
\label{ope}
\end{align}
where $S_F^c(0,x)$ and $S_F^s(x,0)$ are $c$ and $s$ quark propagators in the BFT, $(i z\cdot \tensor{D})^n$ stands for the vertex operators, and ``$\cdots$'' indicates the even higher-order terms.

\begin{figure*}[htb]
\centering
\includegraphics[width=0.95\textwidth]{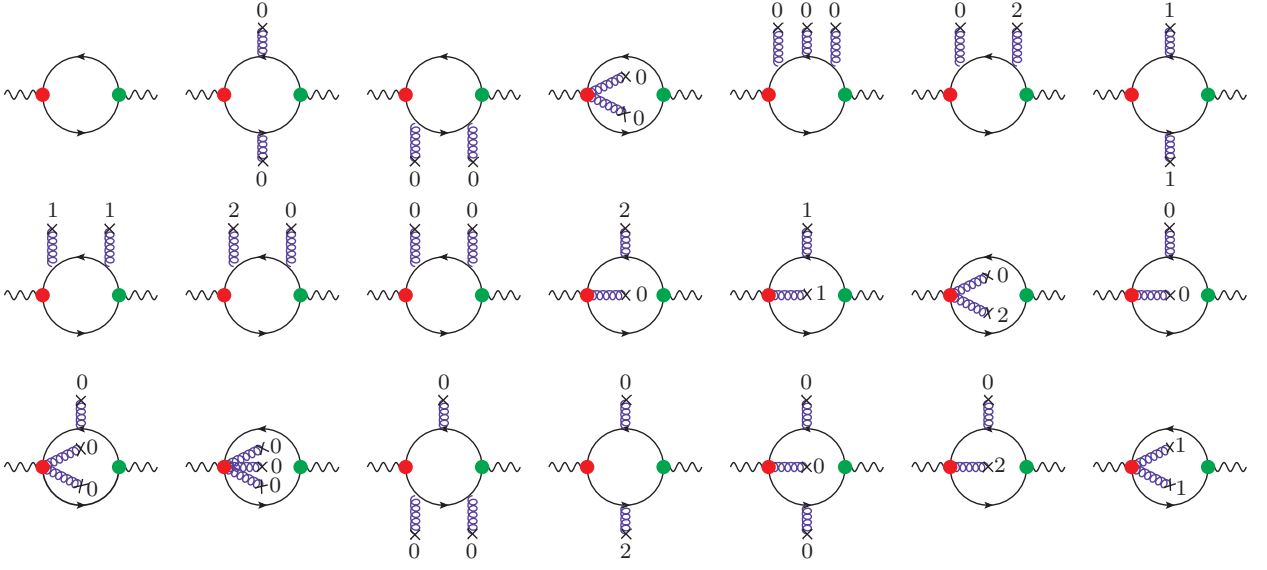}
\caption{Typical Feynman diagrams for the first term of Eq.(\ref{ope}). The left big dot and the right big dot stand for the vertex operators $\DS z \gamma_5 (z\cdot \tensor{D})^n$ and $\DS z \gamma_5$ in the currents $J_n(x)$ and $J^\dagger_0(0)$, respectively. The cross symbol is the gluonic background field.``$n$" indicates $n_{\rm th}$-order covariant derivative.}
\label{feyna}
\end{figure*}

\begin{figure*}[htb]
\centering
\includegraphics[width=0.85\textwidth]{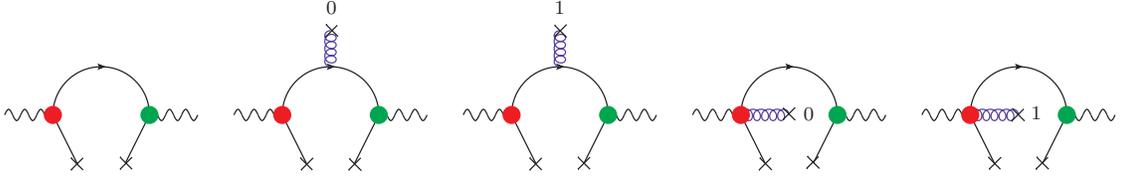}
\caption{Typical Feynman diagrams for the second term of Eq.(\ref{ope}). The left big dot and the right big dot stand for the vertex operators $\not\! z \gamma_5 (z\cdot \tensor{D})^n$ and $\not\! z \gamma_5$ in the currents $J_n(x)$ and $J^\dagger_0(0)$, respectively. The cross symbol attached to the gluon line indicates the tensor of the local gluon background field, and ``$n$" indicates $n_{\rm th}$-order covariant derivative, and the cross symbol attached to the quark line stands for the quark background field.}
\label{feynb}
\end{figure*}

There are totally $40$ Feynman diagrams for the present considered accuracy, e.g. up to dimension-6 operators, the first and second terms in Eq.(\ref{ope}) contain $35$ and $5$ Feynman diagrams, respectively. Typical Feynman diagrams are shown in Figure.~\ref{feyna} and Figure.~\ref{feynb}, other diagrams can be obtained by permutation. In those two figures, the left big dot and the right big dot stand for the vertex operators $\not\! z \gamma_5 (z\cdot \tensor{D})^n$ and $\not\! z \gamma_5$ in the currents $J_n(x)$ and $J^\dagger_0(0)$, respectively; the cross symbol indicates the gluonic background field. There are also cases in which the cross symbol stands for the $s$-quark background field. In deriving the QCD sum rules for the moments, we need to know the propagators and vertex operators under the BFT up to dimension-six operators, and tedious expressions of them can be found in Ref.\cite{BHL_Zhong:2014jla}. Here different from the case of the $D$-meson, the mass effect in the denominator of $s$-quark propagator cannot be ignored. However, considering that the $s$-quark mass is not large, we expand the $s$-quark propagator as a power series over $m_s$ and keep only the first power of $m_s$. In this way, we can use the corresponding calculation technology described in detail in Ref.~\cite{BHL_Zhang:2017rwz} to do the calculation.

Following the standard procedures of QCD sum rules~\cite{Zhong:2021, Hu:2021zmy}, we obtain the sum rules for the moments of $D_s$-meson leading-twist LCDA, i.e.
\begin{align}
&
\frac{\langle \xi^n\rangle _{2;D_s} f_{D_s}^2}{M^2} e^{- m_{D_s}^2/M^2}
\nonumber\\
& \quad = \frac{1}{\pi} \frac{1}{M^2} \int^{s^{D_s}_0}_{t_{\rm min}} ds e^{-\frac{s}{M^2}} {\rm Im} I_{\rm pert}(s) + \hat{\cal B}_{M^2} I_{\langle \bar{s}s\rangle }(-q^2)
\nonumber\\[1.5ex]
& \quad + \hat{\cal B}_{M^2} I_{\langle G^2\rangle }(-q^2) + \hat{\cal B}_{M^2} I_{\langle \bar{s}Gs\rangle }(-q^2) + \hat{\cal B}_{M^2} I_{\langle \bar{s}s\rangle ^2}(-q^2) \nonumber\\[1.5ex]
& \quad + \hat{\cal B}_{M^2} I_{\langle G^3\rangle }(-q^2). \label{Eq:SR}
\end{align}
The analytical expressions of the perturbative and non-perturbative terms are
\begin{widetext}
\begin{align}
&\hspace{-0.5cm}{\rm Im} I_{\rm pert}(s) = \frac{3}{8\pi^2 M^2(n+1)(n+3)} \bigg\{~\bigg[\frac1v\bigg(1+\sqrt{1-\frac{4m_s^2 v^2}{s}}\bigg)-1\bigg]^{n+1} \bigg\{1-\frac{n+1}{2v}\bigg(1+\sqrt{1-\frac{4m_s^2 v^2}{s}}\bigg) \bigg[\frac1v \bigg(1+
\nonumber
\\
&\hspace{1.05cm} \times \sqrt{1-\frac{4m_s^2 v^2}{s}}\bigg)-2\bigg]\bigg\}
-\bigg[\frac1v\bigg(1-\sqrt{1-\frac{4m_s^2 v^2}{s}}\bigg)-1\bigg]^{n+1}\bigg\{1-\frac{n+1}{2v}
\bigg(1-\sqrt{1-\frac{4m_s^2 v^2}{s}}\bigg)
\bigg[\frac1v \bigg(1-
\nonumber
\\
&\hspace{1.05cm}\times \sqrt{1-\frac{4m_s^2 v^2}{s}}\bigg)-2\bigg]\bigg\}\bigg\}
\label{Eq:Perturbative}
\\
&\hspace{-0.5cm}\hat{\cal B}_{M^2} I^{\langle \bar{s}s\rangle }_{2;D_s}(-q^2) = (-1)^n e^{-m_c^2/M^2} \langle  \bar{s}{s} \rangle\bigg[\frac{m_s}{M^4} + \frac{m^2_c m^3_s }{3M^4}+ \frac{(2n+1)m^3_s  }{3M^6}\bigg], \label{Eq:BI_qq}
\\
&\hspace{-0.5cm}\hat{\cal B}_{M^2} I^{\langle G^2\rangle }_{2;D_s}(-q^2) = \frac{\langle \alpha_s G^2\rangle }{12\pi M^4} \left[ 2n(n-1) \mathcal{H}(n-2,1,3,2) + \mathcal{H}(n,0,2,2) - 2m_c^2 \mathcal{H}(n,1,1,3) \right],
\label{Eq:BI_GG}
\\[1.5ex]
&\hspace{-0.5cm}
\hat{\cal B}_{M^2} I_{\langle \bar{s}Gs\rangle }(-q^2) = (-1)^n e^{-m_c^2/M^2}\frac{m_s \langle  g_s\bar{s}\sigma TGs \rangle }{M^6} \left[ -\frac{8n+1}{18} - \frac{2m_c^2}{9M^2} \right],
\label{Eq:BI_qGq}
\\
&\hspace{-0.5cm}
\hat{\cal B}_{M^2} I_{\langle G^3\rangle }(-q^2) = \frac{\langle g_s^3fG^3\rangle }{120\pi^2} ~ \bigg[-10~(n-1)\,n\,(n+1)~\mathcal{H}(n-2,1,4,3) - 30 m^2_c ~ n(n-1)~\mathcal{H}(n-2,1,4,4)-15m^2_c
\nonumber
\\
&\hspace{1.8cm}\times \mathcal{H}(n,1,1,4) - 5 m^2_c \mathcal{H}(n,0,2,4) + 5n m^2_c\mathcal{H}(n-1,1,2,4) +36m^4_c\mathcal{H}(n,1,1,5)\bigg]
\label{Eq:BI_G3}
\\
&\hspace{-0.5cm}\hat{\cal B}_{M^2} I_{\langle \bar{s}s\rangle ^2}(-q^2) = \frac{\langle g_s \bar{s}s\rangle ^2}{2430\pi^2}\bigg[-80n(n+1)\,\,\mathcal{H}(n-2,0,5,3)+120m_c^2n \mathcal{H}(n-1,0,4,4) -60m_c^2\mathcal{H}(n,0,2,4)+180
\nonumber\\
&\hspace{1.88cm}
\times m_c^2 \mathcal{H}(n,0,3,4) + 60(n+1)\, \mathcal{H}(n,0,3,3) + 25 \mathcal{H}(n,0,2,3) - 80n(n+1)\mathcal{H}(n-2,2,3,3) +40n
\nonumber\\[1.2ex]
&\hspace{1.88cm}
\times \mathcal{H}(n-2,1,2,3)-120nm_c^2~\mathcal{H}(n-1,1,3,4)-50n ~\mathcal{H}(n-1,1,2,3) +60nm_c^2~\mathcal{H}(n-1,1,2,4)
\nonumber\\[1.2ex]
&\hspace{1.88cm}
+120n(n-1)m_c^2~\mathcal{H}(n-2,1,4,4) + 40n(n-1)(n+1) \,\mathcal{H}(n-2,1,4,3) - 255\mathcal{H}(n,1,1,3) +45
\nonumber\\
&\hspace{1.88cm}
\times m_c^2 \mathcal{H}(n,1,1,4) -\!144m_c^4\mathcal{H}(n,1,1,5)\bigg]\!+\!\frac{\langle g_s \bar{s}s\rangle ^2}{5832\pi^2M^6 } e^{-m_c^2/M^2} \bigg\{\!\!\!-153[\mathcal{F}_1(n,5,3,2,\infty)\!-\!\mathcal{G}_2(n,5)
\nonumber\\
&\hspace{1.88cm}
+\theta(n-2)~\mathcal{G}_1(n,5)+3\theta(n-1)~ \mathcal{G}_2(n,5)] +30n ~\mathcal{F}_2(n-1,5,3,1,\infty) + 24n\,\mathcal{F}_2(n-2,5,3,1,\infty)
\nonumber\\[1.4ex]
&\hspace{1.88cm} +2 m_c m_s ~\mathcal{F}_2(n,4,4,1,\infty) - 18~ [\,\mathcal{F}_2(n,3,3,1,\infty)  +\mathcal{G}_2(n,3)\,] + 15~[\,\mathcal{F}_2(n,4,3,1,\infty)+\mathcal{G}_2(n,4)]
\nonumber\\
&\hspace{1.88cm}
\times (15+2m_c m_s)+\bigg(\ln \frac{M^2}{\mu^2} - \gamma_E+\frac{3}{2}\bigg)~ [-153 (n+2)\theta(n-1)-3))~-30 ~ (-4\delta_{0n}~-(-1)^n(2n
\nonumber\\
&\hspace{1.88cm}
\times (-1)^n\theta(n-1)+24n+\theta(n-2)~(-1)^n-3(-1)^n]~+\bigg(\ln \frac{M^2}{\mu^2}-\gamma_E+\frac{11}{6}\bigg)~[153(-1)^n+2 m_c
\nonumber\\
&\hspace{1.88cm}
\times m_s (-1)^n]\bigg\}
\label{Eq:BI_ss2}
\end{align}
\end{widetext}
with $v = s/(s-m^2_c+m^2_s)$. Here the functions $\mathcal{F}_{1,2}(n,a,b,l_{\rm min},l_{\rm max}), \mathcal{G}_{1,2}(n,a), \mathcal{H}(n,a,b,c)$ and Borel transformations which are collected in the Appendix~\ref{Sec:AppendixA}.

\subsection{The light-cone harmonic oscillator model for the $D_s$-meson leading-twist LCDA $\phi_{2;D_s}$}

Based on the BHL-description~\cite{BHL1,BHL2,BHL3}, similar to the case of $D$-meson leading-twist LCDA~\cite{BHL_Zhang:2017rwz}, we construct a light-cone harmonic oscillator model of the $D_s$-meson leading-twist wavefunction $\Psi_{2;D_s}(x,\mathbf{k}_\perp)$ as
\begin{eqnarray}
\Psi_{2;D_s}(x,\mathbf{k}_\perp) = \chi_{2;D_s}(x,\mathbf{k}_\perp) \Psi_{2;D_s}^R(x,\mathbf{k}_\perp),
\label{wf}
\end{eqnarray}
where $\mathbf{k}_\perp$ is the transverse momentum, $\chi_{2;D_s}(x,\mathbf{k}_\perp)$ is the spin-space wavefunction and $\Psi_{2;D_s}^R(x,\mathbf{k}_\perp)$ indicates the spatial wavefunction. The spin-space wavefunction $\chi_{2;D_s}(x,\mathbf{k}_\perp)$ reads~\cite{Huang:1994dy}
\begin{eqnarray}
\chi_{2;D_s}(x,\mathbf{k}_\perp) = \frac{\hat{m}_c x + \hat{m}_s(1-x)}{\sqrt{\mathbf{k}_\perp^2 + \left[ \hat{m}_c x + \hat{m}_s(1-x) \right]^2}},
\label{chiwf}
\end{eqnarray}
where $\hat{m}_c$ and $\hat{m}_s$ are constituent quark masses of $D_s$, and we adopt $\hat{m}_c = 1.5 \rm GeV$ and $\hat{m}_s = 0.5 \rm GeV$. The spatial wavefunction takes the form
\begin{align}
&\Psi_{2;D_s}^R(x,\mathbf{k}_\perp) = A_{D_s} \varphi_{2;D_s}(x) \nonumber\\
&\qquad \times \exp \left[ -\frac{1}{\beta_{D_s}^2} \left( \frac{\mathbf{k}_\perp^2 + \hat{m}_c^2}{1-x} + \frac{\mathbf{k}_\perp^2 + \hat{m}_s^2}{x} \right) \right],
\label{psirwf}
\end{align}
where $A_{D_s}$ is the normalization constant, $\beta_{D_s}$ is the harmonious parameter that dominates the wavefunction's transverse distribution, and function $\varphi_{2;D_s}(x)$ dominates the wavefunction's longitudinal distribution. $\varphi_{2;D_s}(x)$ can be taken as the first few terms of the Gegenbauer series, here we take
\begin{eqnarray}
\varphi_{2;D_s}(x) = 1 + \sum^4_{n=1} B_n^{D_s} C^{3/2}_n(2x-1).
\end{eqnarray}
By using the relationship between the $D_s$-meson leading-twist wavefunction, one can get its LCDA at the scale $\mu_0$,
\begin{eqnarray}
\phi_{2;D_s}(x,\mu_0) = \frac{2\sqrt{6}}{f_{D_s}} \int_{|\mathbf{k}_\perp|^2\leq\mu_0^2} \frac{d^2\mathbf{k}_\perp}{16\pi^3} \Psi_{2;D_s}(x,\mathbf{k}_\perp),
\end{eqnarray}
which, after integrating over the transverse momentum $\mathbf{k}_\perp$, becomes
\begin{eqnarray}
\phi_{2;D_s}(x,\mu_0) &=& \frac{\sqrt{6} A_{D_s} \beta_{D_s}^2}{\pi^2 f_{D_s}} x(1-x) \varphi_{2;D_s}(x) \nonumber\\
&\times& \exp \left[ - \frac{\hat{m}_c^2x + \hat{m}_s^2(1-x)}{8\beta_{D_s}^2 x(1-x)} \right] \nonumber\\
&\times& \left\{ 1 - \exp \left[ -\frac{\mu^2_0}{8\beta_{D_s}^2 x(1-x)} \right] \right\},
\label{phi}
\end{eqnarray}
where $\mu_0 \sim \Lambda_{\rm QCD}$ is the factorization scale. Because $\hat{m}_c \gg \Lambda_{\rm QCD}$, the spin-space wavefunction $\chi_{D_s} \to 1$. The above model (\ref{wf}, \ref{phi}) is for $D_s^-$-meson. The leading-twist wavefunction and the LCDA for $D_s^+$-meson can be obtained by replacing $x$ with $(1-x)$ in Eqs.(\ref{wf}, \ref{phi}).

The model parameters $A_{D_s}$, $B_n^{D_s}$ and $\beta_{D_s}$ are scale dependent, their values at an initial scale $\mu_0$ can be determined by reasonable constraints, and their values at any other scale $\mu$ can be derived via the evolution equation~\cite{Lepage:1980fj}. More explicitly, we shall adopt the following constraints to fix the parameters:
\begin{itemize}
\item The normalization condition,
\begin{eqnarray}
\int^1_0 dx \phi_{2;D_s}(x,\mu_0) = 1. \label{NC}
\end{eqnarray}

\item The probability of finding the leading Fock-state $|\bar{c}s\rangle $ in $D_s$-meson Fock-state expansion,
\begin{eqnarray}
P_{D_s} &=& \frac{A_{D_s}^2 \beta_{D_s}^2}{4\pi^2} x(1-x) \varphi_{D_s}^2(x) \nonumber\\
&\times& \exp \left[ -\frac{m_c^2x + m_s^2 (1-x)}{4\beta^2_{D_s} x(1-x)} \right].   \label{P}
\end{eqnarray}
We will take $P_{D_s} \simeq 0.8$ ~\cite{MOLELIV_Guo:1991eb} in subsequent calculations.

\item The Gegenbauer moments of $\phi_{2;D_s}(x,\mu_0)$ can be derived via the following formula,
\begin{eqnarray}
a_n^{D_s}(\mu_0) = \frac{\int^1_0 dx \phi_{2;D_s}(x,\mu_0)C_n^{3/2}(2x-1)}{\int^1_0 dx 6x(1-x) [C_n^{3/2}(2x-1)]^2}, \label{an}
\end{eqnarray}
and the $\phi_{2;D_s}(x,\mu_0)$ moments are defined as
\begin{eqnarray}
\langle \xi^n\rangle _{2;D_s}|_{\mu_0} = \int^1_0 dx (2x-1)^n \phi_{2;D_s}(x,\mu_0).
\end{eqnarray}
The values of the moments $\langle \xi^n\rangle _{2;D_s}$ and the Gegenbauer moments $a_n^{D_s}$ at the scale $2$ GeV will be given in next subsection.

\end{itemize}

\section{numerical analysis} \label{Sec:III}

\subsection{Input parameters}

To do the numerical analysis on the moments of $D_s$-meson leading-twist LCDA, we take the $D_s$-meson mass $m_{D_s}=1.968 \pm 0.00007 ~{\rm GeV}$, the $c$-quark current-quark mass $\bar{m}_{c}(\bar{m}_c) = 1.275 \pm 0.02 ~{\rm GeV}$, the $s$-quark mass $m_s(2~{\rm GeV}) = 0.093^{+0.011}_{-0.005} ~{\rm GeV}$ and the decay constant of $D_s$-meson $f_{D_s} = 0.256 \pm 0.0042 \rm MeV$~\cite{Zyla:2020zbs}. For the gluon condensates, we take $\langle \alpha_sG^2\rangle  = 0.038 \pm 0.011 ~{\rm GeV}^4$ and $\langle g_s^3fG^3\rangle  = 0.045 ~{\rm GeV}^6$~\cite{SRREV_Colangelo:2000dp}. For the remaining vacuum condensates, we adopt $\langle \bar{s}s\rangle  = \kappa \langle \bar{q}q\rangle $, $\langle g_s\bar{s}\sigma TGs\rangle  = \kappa \langle g_s\bar{q}\sigma TGq\rangle $, and $\langle g_s\bar{s}s\rangle ^2 = \kappa^2 \langle g_s\bar{q}q\rangle ^2$, where $\kappa = 0.74\pm0.03$~\cite{Narison:2014wqa}, $\langle \bar{q}q\rangle  = \left( -2.417^{+0.227}_{-0.114} \right) \times 10^{-2} ~{\rm GeV}^2$, $\langle g_s\bar{q}\sigma TGq\rangle  = \left( -1.934^{+0.188}_{-0.103} \right) \times 10^{-2} ~{\rm GeV}^5$ and $\langle g_s\bar{q}q\rangle ^2 = \left( 2.082^{+0.743}_{-0.697} \right) \times 10^{-3} ~{\rm GeV}^6$ at $\mu = 2 ~{\rm GeV}$~\cite{Zhong:2021}. The scale evolution equations of those inputs are~\cite{Zhong:2021,RGE_Yang:1993bp, RGE_Hwang:1994vp}
\begin{align}
&\bar{m}_c(\mu) = \bar{m}_c(\bar{m}_c) \left[ \frac{\alpha_s(\mu)}{\alpha_s(\bar{m}_c)} \right]^{4/\beta_0},
\nonumber\\
&\bar{m}_s(\mu) = \bar{m}_s(2~{\rm GeV}) \left[ \frac{\alpha_s(\mu)}{\alpha_s(2~{\rm GeV})} \right]^{4/\beta_0},
\nonumber\\
&\langle \bar{q}q\rangle (\mu) = \langle \bar{q}q\rangle (2~{\rm GeV}) \left[ \frac{\alpha_s(\mu)}{\alpha_s(2~{\rm GeV})} \right]^{-4/\beta_0},
\nonumber\\
&\langle g_s\bar{q}\sigma TGq\rangle (\mu) = \langle g_s\bar{q}\sigma TGq\rangle (2~{\rm GeV}) \left[ \frac{\alpha_s(\mu)}{\alpha_s(2~{\rm GeV})} \right]^{-{2}/{(3\beta_0)}},
\nonumber\\
& \langle g_s\bar{q}q\rangle ^2(\mu) = \langle g_s\bar{q}q\rangle ^2 \left[ \frac{\alpha_s(\mu)}{\alpha_s(2~{\rm GeV})} \right]^{-4/\beta_0},
\nonumber\\
&\langle \alpha_sG^2\rangle (\mu) = \langle \alpha_sG^2\rangle (\mu_0), \nonumber\\
&\langle g_s^3fG^3\rangle (\mu) = \langle g_s^3fG^3\rangle (\mu_0),
\end{align}
where $\beta_0 = 11-2n_f/3$ with $n_f$ being the active quark flavors. In the following numerical calculation of the moments $\langle \xi^n\rangle _{2;D_s}|_{\mu}$, the scale $\mu$ will be set as the Borel parameter as usual, i.e., $\mu=M$. For the continuous threshold $s_0^{D_s}$, it is usually taken as the squared mass of the $D_s$-meson's first exciting state, and we take $s_0^{D_s} \simeq 6.5 ~\textrm{GeV}^2$.

\subsection{The moments $\langle \xi^n\rangle _{2;D_s}$ from QCD sum rules}

\begin{table}[htb]
\caption{The determined Borel windows and the corresponding $D_s$-meson leading-twist LCDA moments $\langle \xi^n\rangle _{2;D_s}$ with $ n=(1,2,3,4)$. All input parameters are set to be their central values. $\mu=M$. }
\begin{tabular}{ c  c  c }
\hline
~$n$~ & ~~~~~~~~~~~~~~~~~~~$M^2$~~~~~~~~~~~~~~~~~~~ & ~$\langle \xi^n\rangle _{2;D_s}$~ \\
\hline
~$1$~ & ~$[1.517,5.840]$~ & ~$[-0.304,-0.263]$~ \\
~$2$~ & ~$[1.265,4.164]$~ & ~$[+0.168,+0.193]$~ \\
~$3$~ & ~$[2.162,7.185]$~ & ~$[-0.107,-0.104]$~ \\
~$4$~ & ~$[1.928,5.524]$~ & ~$[+0.069,+0.077]$~ \\
\hline
\end{tabular}
\label{tbw}
\end{table}

\begin{figure}[htb]
\centering
\includegraphics[width=0.42\textwidth]{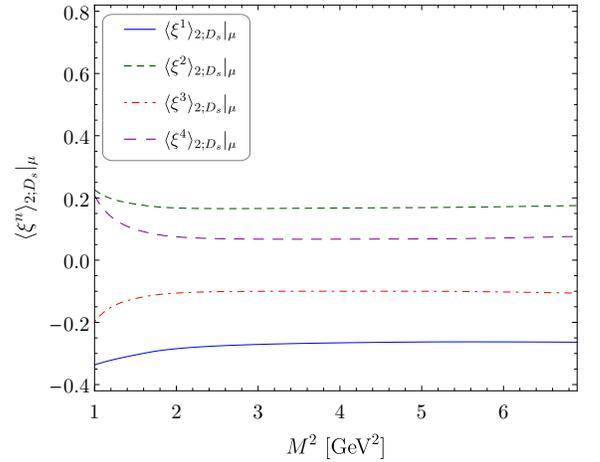}
\caption{The $D_s$-meson leading-twist LCDA moments $\langle \xi^n\rangle _{2;D_s}$ at the scale $\mu=M$ with $n=(1,\cdots,4)$ versus the Borel parameter $M^2$, where all input parameters are set to be their central values. }
\label{fxinM2}
\end{figure}

To get the numerical value of moments $\langle \xi^n\rangle _{2;D_s}$ of $\phi_{2;D_s}(x,\mu)$, one need to fix the Borel window $M^2$ which is introduced to depress the contributions from both the continuum states and the highest dimensional condensates. Usually, the continuum contribution and the dimension-six condensate contribution are taken to be less than $30\%$ and $10\%$ respectively, while the value of $\langle \xi^n\rangle _{2;D_s}$ is required to be as stable as possible in the allowed Borel window. In this paper, the continuum state contribution for $\langle \xi^n\rangle _{2;D_s}|_\mu$ with $n =(1,2,3,4)$ is required to be less than $20\%, 25\%, 10\%, 30\%$, respectively, and each of the dimension-six condensates contributions is no more than $1\%$. The determined Borel windows and the corresponding $D_s$-meson leading-twist LCDA moments $\langle \xi^n\rangle _{2;D_s}$ at the scale $\mu=2$ GeV with $n=(1,\cdots,4)$ are presented in Table~\ref{tbw}, where all input parameters are taken to be their central values. We present the $D_s$-meson leading-twist LCDA moments $\langle \xi^n\rangle _{2;D_s}$ with $n=(1,\cdots,4)$ at $\mu=$ 2GeV versus $M^2$ in Fig.~\ref{fxinM2}. To be consistent with Table~\ref{tbw}, those moments are stable over the allowable Borel windows.

If setting $\mu=2$ GeV, by taking all uncertainty sources into consideration, we obtain
\begin{eqnarray}
\langle \xi^1\rangle _{2;D_s} |_{\mu = 2 \rm GeV} &=& -0.261^{+0.020}_{-0.020}, \label{xi11}\\
\langle \xi^2\rangle _{2;D_s} |_{\mu = 2 \rm GeV} &=& +0.184^{+0.012}_{-0.012}, \label{xi22} \\
\langle \xi^3\rangle _{2;D_s} |_{\mu = 2 \rm GeV} &=& -0.111^{+0.007}_{-0.012}, \label{xi33} \\
\langle \xi^4\rangle _{2;D_s} |_{\mu = 2 \rm GeV} &=& +0.075^{+0.005}_{-0.005}, \label{xi44}
\label{nxin}
\end{eqnarray}
where the errors are squared averages of all the mentioned error sources.

\subsection{Determination of the model parameters of $\phi_{2;D_s}$}

\begin{table*}[htb]
\caption{Typical $D_s$-meson leading-twist LCDA model parameters at scale $\mu =2~{\rm GeV}$. The first line stand for the central value, the secone/third lines mean the upper/lower limit for the LCDA.}
\begin{tabular}{ l l l l l  c c c c c c}
\hline
&~~$a_1^{D_s}(\mu)$& ~~$a_2^{D_s}(\mu)$ & ~$a_3^{D_s}(\mu)$ & ~~$a_4^{D_s}(\mu)$& $A_{D_s}({\rm GeV}^{-1})$ & $B^{D_s}_1$ & $B^{D_s}_2$& $B^{D_s}_3$ & $B^{D_s}_4$ & $\beta_{D_s}({\rm GeV})$ \\
\hline
Central value &$-0.436$& $-0.047$ & $0.004$ & $-0.004$& $2.760$ & $-0.313$ & $-0.185$& $0.083$ & $0.008$ & $4.521$
\\
Upper&$-0.436^{+0.033}$& $-0.047_{-0.035}$ & $0.004^{+0.010}$ & $-0.004_{-0.026}$& $2.802$ & $-0.290$ & $-0.198$& $0.079$ & $0.001$ & $4.484$
\\
Lower&$-0.436_{-0.033}$& $-0.047^{+0.035}$ & $0.004_{-0.020}$ & $-0.004^{+0.025}$& $2.717$ & $-0.334$ & $-0.173$& $0.081$ & $0.014$ & $4.567$ \\
\hline
\end{tabular}
\label{DAparameter}
\end{table*}

According to the constraints of $D_s$-meson leading-twist LCDA $\phi_{2;D_s}(x,\mu)$, i.e., Eqs.(\ref{NC}-\ref{an}), we need to know the Gegenbauer moments $a_n^{D_s}(\mu)$ to fix the parameters $A_{D_s}$, $B_n^{D_s}$ and $\beta_{D_s}$. The Gegenbauer moments $a_n^{D_s}(\mu)$ using their relations to the LCDA moments $\langle \xi^n\rangle _{2;D_s}|_\mu$~\cite{BHL_Zhong:2014fma}, we obtain
\begin{eqnarray}
a_1^{D_s}(2~ \rm GeV) &=& -0.436^{+0.033}_{-0.033}, \\
a_2^{D_s}(2~ \rm GeV) &=& -0.047^{+0.035}_{-0.035}, \\
a_3^{D_s}(2~ \rm GeV) &=& +0.004^{+0.010}_{-0.020}, \\
a_4^{D_s}(2~ \rm GeV) &=& -0.004^{+0.025}_{-0.026}.
\label{anD}
\end{eqnarray}
We present all the determined input parameters at the scale $\mu =2~{\rm GeV}$ in Table~\ref{DAparameter}. The accuracy of $\phi_{2;D_s}(x,\mu)$ is dominated by the magnitudes of the Gegenbauer moments $a_n^{D_s}(\mu)$. As we have pointed out in Ref.~\cite{BHL_Zhang:2017rwz, BHL_Zhong:2018exo}, the Gegenbauer moments $a_n^{D_s}(\mu)$ are correlated to each other and can not be changed independently within their own error regions. Then Table \ref{DAparameter} associates the uncertainty of $\phi_{2;D_s}(x,\mu)$ with the error of Gegenbauer moments $a_n^{D_s}(\mu)$, which facilitates our further discussion on the impact of $\phi_{2;D_s}(x,\mu)$ as an input parameter to the $B_s\to D_s$ decay.

\begin{figure}[htb]
\centering
\includegraphics[width=0.42\textwidth]{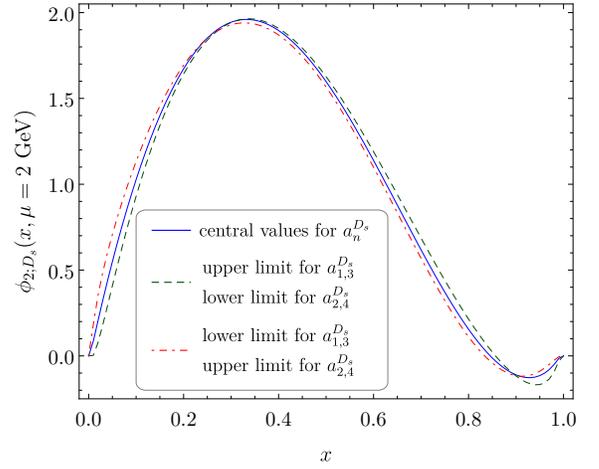}
\caption{The $D_s$-meson leading-twist LCDA $\phi_{2;D_s}(x,\mu)$ with the parameter values exhibited in Table \ref{DAparameter}. }
\label{fDAan}
\end{figure}

\begin{figure}[htb]
\centering
\includegraphics[width=0.42\textwidth]{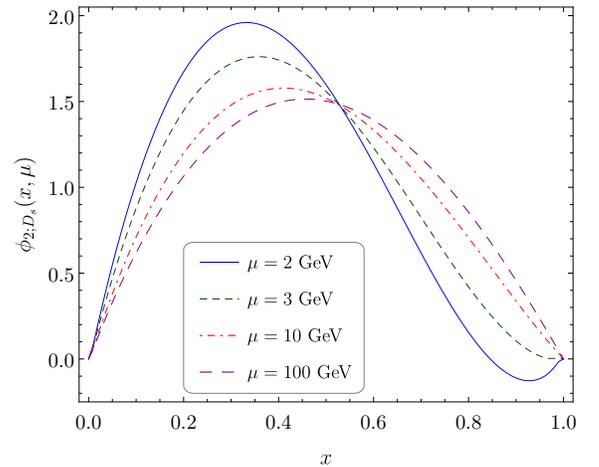}
\caption{The $D_s$-meson leading-twist LCDA $\phi_{2;D_s}(x,\mu)$ at different scales, where the solid, the dashed, the dotted and the dash-dotted lines are for $\mu = 2, 3, 10, 100$ GeV, respectively.}
\label{fDAEverlution}
\end{figure}

Figure.~\ref{fDAan} shows the $D_s$-meson leading-twist LCDA $\phi_{2;D_s}(x,\mu)$ with typical values of the input parameters exhibited in Table \ref{DAparameter}. The solid, the dash-dotted and the dashed lines are for the parameters exhibited in second, third and forth lines of Table \ref{DAparameter}. Our model of $\phi_{2;D_s}(x,\mu)$ prefers a broader behavior in low $x$-region. It has a peak around $x\sim 0.35$. Figure.~\ref{fDAEverlution} shows the $D_s$-meson leading-twist LCDA $\phi_{2;D_s}(x,\mu)$ at different scales, where the solid, the dashed, the dotted and the dash-dotted lines are for the scales $\mu = 2, 3, 10, 100$ GeV, respectively. It shows that with the increment of $\mu$, $\phi_{2;D_s}(x,\mu)$ becomes broader and broader and becomes more symmetric, e.g. the peak moves closer to $x=0.5$. When $\mu\to\infty$, $\phi_{2;D_s}(x,\mu)$ tends to the known asymptotic form, i.e. $\phi_{2;D_s}(x,\mu\to\infty)=6x(1-x)$.

\subsection{Numerical results of $B_s\to D_s$ TFF and its applications}

Our inputs for the $B_s\to D_s$ TFF $f^{B_s \to D_s}_+(q^2)$ are~\cite{Zyla:2020zbs}
\begin{align}
&m_{\bar B_s^0}= 5.36688 \pm 0.00017 ~{\rm GeV}, \nonumber\\
&\bar{m}_b (\bar{m}_b)=4.18^{+0.04}_{-0.03}~{\rm GeV}, \nonumber\\
&f_{B_s}=266 \pm 19~ {\rm MeV}. \nonumber
\end{align}

\begin{figure}[htb]
\centering
\includegraphics[width=0.42\textwidth]{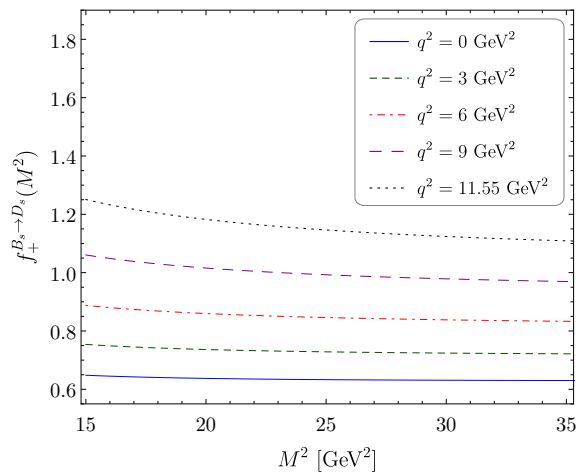}
\caption{The TFF $f^{B_s\to D_s}_+(q^2)$ for some typical $q^2$ values versus the Borel parameter $M^2$.}
\label{ftffm}
\end{figure}

\begin{table}[htb]
\caption{The parameters $a$ and $b$ for the TFF extrapolation. The lowest, middle and the highest TFFs are adopted for such a determination.}
\begin{tabular}{ c  c  c }
\hline
~$f^{B_s \to D_s}_+(0)$~ & ~~~~~~~~~~~~~~~~~~~$a$~~~~~~~~~~~~~~~~~~~ & ~$b$~ \\
\hline
~$0.639$~ & ~$1.350$~ & ~$0.479$~ \\
~$0.583$~ & ~$1.345$~ & ~$0.531$~ \\
~$0.714$~ & ~$1.320$~ & ~$0.443$~ \\
\hline
\end{tabular}
\label{extra}
\end{table}

There are still two parameters to be fixed, the continuum threshold $s_0^{B_s}$ and the Borel window $M^2$. We set $s_0^{B_s} = 38 \pm 1 ~\textrm{GeV}^2$ and $M^2 = (20-30) ~\textrm{GeV}^2$ with the scale $\mu \simeq 3 ~{\rm GeV}$ which is close to $\sqrt{m_{{B}_s}^2-m_b^2}$. Such a choice makes the TFF $f^{B_s\to D_s}_+(q^2)$ be stable within the allowable Borel window as can be seen from Fig.~\ref{ftffm}. In large recoil point $q^2=0$, we obtain
\begin{eqnarray}
f^{B \to D}_{+}(0) &=& 0.639^{+0.056}_{-0.009}|_{\phi_{2;D_s}}\ ^{+0.005}_{-0.013}|_{M^2}\ ^{+0.014}_{-0.015}|_{s_0^{B_s}}\nonumber\\
 &&^{+0.043}_{-0.049}|_{f_{B_s}} \pm 0.010|_{f_{D_s}}\ ^{+0.018}_{-0.012}|_{m_b},
\end{eqnarray}
and in zero recoil region $q^2=q^2_{\rm max}$, we obtain
\begin{eqnarray}
f^{B \to D}_{+}(q^2_{\rm max}) &=& 1.189 \pm 0.125,
\end{eqnarray}
where all the uncertainties have been added up in quadrature, and the errors from $\phi_{2;D_s}(x,\mu)$ and $f_{B_s}$ dominate the uncertainties. It agrees with the lattice QCD predictions within errors, $f^{B_s\to D_s}_+(0) =0.656(31)$~\cite{Monahan:2017uby} and $f^{B_s\to D_s}_+(0) = 0.666(12)$~\cite{McLean:2019qcx}.

Fig.~\ref{ftffm} also shows that for larger $q^2$-values, the TFF will show sizable dependence on $M^2$, which agrees with the convention that the LCSR approach cannot be applied for very large $q^2$-value. We adopt the TFF $f^{B_s \to D_s}_{+}(q^2)$ within the region of $[0, 9{\rm GeV}^2]$ as a basis to extrapolate it to all physical $q^2$-value. For the purpose, we adopt the double-pole-extrapolation method~\cite{Wang:2008xt} to do the extrapolation, i.e.
\begin{eqnarray}
f^{B_s\to D_s}_+(q^2)=\frac{f^{B_s\to D_s}_+(0)}{1-a(q^2/m^2_{B_s})+b(q^2/m^2_{B_s})^2}.
\end{eqnarray}
We put the fitted parameters in Table \ref{extra}.

\begin{figure}[htb]
\centering
\includegraphics[width=0.41\textwidth]{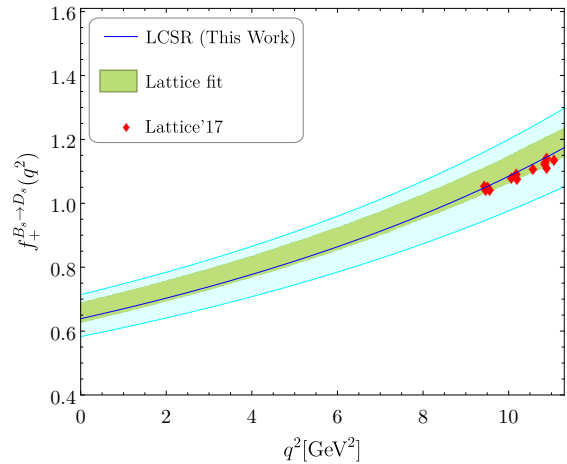}
\caption{The extrapolated LCSR prediction for the TFF $f^{B_s\to D_s}_+(q^2)$, where the lighter shaded band is the squared average of those from all the mentioned error sources. The Lattice QCD prediction and its extrapolated results given in year 2017~\cite{Monahan:2017uby} have also been presented as a comparison, the thicker shaded band shows its uncertainty.}
\label{ftffq}
\end{figure}

The extrapolated results are presented in Fig.~\ref{ftffq}, where the solid line is the central value of $f^{B_s\to D_s}_+(q^2)$, and the lighter shaded band shows its theoretical uncertainty, in which the uncertainties from all the mentioned error sources, such as $\phi_{2;D_s}(x,\mu)$, $s_0^{B_s}$, $f_{B_s}$, $f_{D_s}$, $m_b$ and etc., have been added up in quadrature. As a comparison, the Lattice QCD predictions for large $q^2$-points and its extrapolation to all $q^2$-region have also been presented and the thicker shaded band represents the errors~\cite{Monahan:2017uby}. Our results agree well with the Lattice QCD predictions, especially the arising trends over the changes of $q^2$.

\begin{table}[htb]
\caption{A comparison of $|V_{cb}|$ under various approaches and the experimental measured values.}
\begin{tabular}{l l  }
\hline
References~~~~~~~~~~~~~~~~~~~~~~~~~~~~~~~~~ & $|V_{cb}| \times 10^{-3}$ \\
\hline
This work                                                    & $40.003^{+4.929}_{-4.075}$ \\
LHCb(CLN)~\cite{Aaij:2020hsi}                    & $41.4 \pm 0.6 \pm 0.9 \pm 1.2$ \\
LHCb(BGL)~\cite{Aaij:2020hsi}                    & $42.3 \pm 0.8 \pm 0.9 \pm 1.2$ \\
HPQCD ~\cite{LQCD_Na:2015kha}               & $39.6 \pm 1.7 \pm 0.2 $ \\
PDG~\cite{Zyla:2020zbs}                            & $41.0 \pm 1.4 $ \\
BaBar ~\cite{Dey:2019bgc}                         & $38.36 \pm 0.9 $ \\
BELLE(CLN+LQCD)~\cite{Waheed:2018djm} & $38.4 \pm 0.2 \pm 0.6 \pm 0.6$ \\
BELLE(BGL+LQCD)~\cite{Waheed:2018djm} & $38.3 \pm 0.3 \pm 0.7 \pm 0.6$ \\
LQCD~\cite{Harrison:2017fmw}                  & $41.3 \pm 2.2 $ \\
\hline
\end{tabular}
\label{Tab:Vcb}
\end{table}

As applications, we adopt the LCSR prediction for the TFF to make a prediction on the CKM matrix element $|V_{cb}|$ and the branching ratio $\mathcal{B}(B_s\to D_s \ell\bar{\nu}_\ell)$.

The TFF at the zero-recoil point, $ f^{B_s\to D_s}_+(q^2_{\rm max})$, is
often quoted as
\begin{equation}
\mathcal{G}(1)=\frac{2\sqrt{m_{B_s}m_{D_s}}}{m_{B_s}+m_{D_s}} \times f^{B_s\to D_s}_+(q^2_{\rm max}).
\end{equation}
Using the averaged value given by the Babar collaboration via the measurements on the semi-leptonic decay $\bar{B} \to D\ell\bar\nu_\ell$~\cite{Aubert:2008yv, Aubert:2009ac}, $\eta_{\rm ew} \mathcal{G}(1) |V_{cb}|=(42.65 \pm 1.53) \times 10^{-3}$, one obtain $|V_{cb}|=\left(40.003^{+4.929}_{-4.075}\right)\times 10^{-3}$. In Table~\ref{Tab:Vcb}, we present a comparison of $|V_{cb}|$ with the LHCb measured values under CLN and BGL approaches~\cite{Aaij:2020hsi}, the HPQCD prediction~\cite{LQCD_Na:2015kha}, the PDG averaged value~\cite{Zyla:2020zbs}, the BaBar measured value~\cite{Dey:2019bgc}, the BELLE measured values under CLN+LQCD and BGL+LQCD approaches~\cite{Waheed:2018djm} and the Lattice QCD prediction~\cite{Harrison:2017fmw}.

\begin{figure}[htb]
\centering
\includegraphics[width=0.42\textwidth]{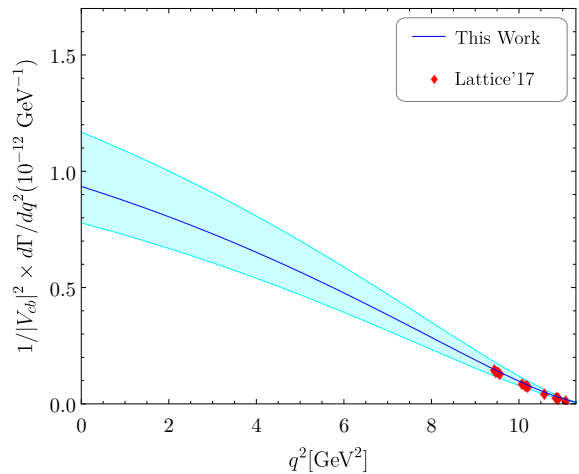}
\caption{The differential decay width $1/|V_{cb}|^2\times d\Gamma/dq^2$ with $\ell = (e,\mu)$. As a comparison, we also present the Lattice QCD predictions in large $q^2$ points~\cite{Monahan:2017uby}.}
\label{Fig:dG}
\end{figure}

We adopt the extrapolated TFF $f^{B_s \to D_s}_{+}(q^2)$ to calculate the branching ratio $\mathcal{B}(B_s\to D_s \ell\bar{\nu}_\ell)$, which can be derived by using the following formula
\begin{align}
\mathcal{B}(B_s\to D_s \ell\bar{\nu}_\ell) = \tau_{B_s} \int^{(m_{B_s} - m_{D_s})^2}_{0} \!\!\!dq^2 \frac{d\Gamma(B_s\to D_s\ell\bar{\nu}_\ell)}{dq^2},
\label{Bra_Rat}
\end{align}
where the differential decay width is
\begin{align}
\frac{d\Gamma(B_s\to D_s \ell\bar{\nu}_\ell) }{dq^2} = \frac{G_F^2 |V_{\rm cb}|^2}{192\pi^3 m_{B_s}^3} \lambda^{3/2}(q^2) |f_+^{B_s\to D_s}(q^2)|^2,
\label{Dif_Dec_Wid}
\end{align}
where $G_F = 1.1663787(6) \times 10^{-5} \rm{GeV}^{-2}$, and the phase-space factor $\lambda(q^2) = (m_{B_s}^2 + m_{D_s}^2 - q^2)^2 - 4m_{B_s}^2 m_{D_s}^2$. We present the differential decay width $1/|V_{cb}|^2\times d\Gamma/dq^2$ in Fig.~\ref{Fig:dG}. After considering the $B_s$-meson lifetime $\tau_{B_s} = (1.510 \pm 0.004) \times 10^{-12} $s ~\cite{Zyla:2020zbs}, we obtain
\begin{eqnarray}
\mathcal{B}(\bar{B}_s^0\to D_s^+\ell\nu_\ell) &=& \left(2.033^{+0.350}_{-0.488}\right) \times 10^{-2}.
\label{Br}
\end{eqnarray}

\section{summary} \label{Sec:IV}

In this work, we have made a detailed study on the $D_s$-meson leading-twist LCDA $\phi_{2;D_s}$. Its moments have been calculated by using the QCD sum rules within the framework of BFT, and its first four moments have been given in Eqs.(\ref{xi11}, \ref{xi22}, \ref{xi33}, \ref{xi44}), which then result in the Gegenbauer moments $a_1^{D_s}(2{\rm GeV}) = -0.436^{+0.033}_{-0.033}$, $a_2^{D_s}(2{\rm GeV}) = -0.047^{+0.035}_{-0.035}$, $a_3^{D_s}(2{\rm GeV}) = 0.004^{+0.01}_{-0.02}$ and $a_4^{D_s}(2{\rm GeV})= -0.004_{+0.025}^{-0.026}$. Based on the BHL-prescription, we have constructed a new model for $\phi_{2;D_s}$, whose behavior is constrained by the normalization condition, the probability of finding the leading Fock-state $|\bar{c}s\rangle $ in $D_s$-meson Fock-state expansion, and the known Gegenbauer moments. As the key input for studying the high-energy processes involving $D_s$-meson, our suggested $\phi_{2;D_s}$ shall be of great importance.

Using the present model of $\phi_{2;D_s}$, we calculate the $B_s\to D_s$ TFF $f^{B_s\to D_s}_+(q^2)$ within the QCD LCSR approach by adopting a chiral current correlator, in which the leading-twist terms dominant over the LCSR. At the large recoil region, we obtain $f^{B_s\to D_s}_+(0) = 0.639^{+0.075}_{-0.056}$. By using the extrapolated TFF with the double-pole-extrapolation method, we obtain $\mathcal{B}(\overline{B}_s^0\to {D_s}^+\ell\nu_\ell) = \left(2.033^{+0.350}_{-0.488} \right)\times 10^{-2}$ and the CKM element $|V_{cb}|=(40.00_{-4.075}^{+4.929})\times 10^{-3}$, which is consistent with the various measurements within reasonable errors.  \\

{\bf Acknowledgments}: This work was supported in part by the National Natural Science Foundation of China under Grant No.11625520, No.11765007, No.11947406 and No.12047564, the Graduate Research and Innovation Foundation of Chongqing, China (Grant No.ydstd1912), the Project of Guizhou Provincial Department of Science and Technology under Grant No.KY[2019]1171, the Project of Guizhou Provincial Department of Education under Grant No.KY[2021]030 and No.KY[2021]003, the China Postdoctoral Science Foundation under Grant Nos.2019TQ0329, 2020M670476, the Chongqing Graduate Research and Innovation Foundation under Grant No.2020CQJQY-Z003, and the Project of Guizhou Minzu University under Grant No. GZMU[2019]YB19.

\appendix

\section*{Appendix: Useful functions for calculating the moments of $\phi_{2;D_s}$} \label{Sec:AppendixA}

The functions $\mathcal{F}_{1,2}(n,a,b,l_{\rm min},l_{\rm max}), \mathcal{G}_{1,2}(n,a), \mathcal{H}(n,a,b,c)$ used in the sum rules Eqs.~\eqref{Eq:Perturbative}-\eqref{Eq:BI_ss2}
\begin{widetext}
\begin{align}
\mathcal{F}_1(n,a,b,l_{\rm min},l_{\rm max}) &= \sum_{k=0}^n \frac{(-1)^k n! \Gamma(k+a)}{k!(n-k)!} \sum_{l=l_{\rm min}}^{l_{\rm max}} \frac{\Gamma(l+b) \Gamma(n-1-k+l)}{\Gamma(n-1+l+a)}
\nonumber\\
&\times \sum^l_{i=0} \frac{1}{i! (l-i)! (l-1-i+b)!} \left( -\frac{m_c^2}{M^2} \right)^{l-i},
\label{F1}
\\
\mathcal{F}_2(n,a,b,l_{\rm min},l_{\rm max}) &= \sum_{k=0}^n \frac{(-1)^k n! \Gamma(k+a)}{k!(n-k)!} \sum_{l=l_{\rm min}}^{l_{\rm max}} \frac{\Gamma(l+b) \Gamma(n-k+l)}{\Gamma(n+l+a)}
\nonumber\\
&\times \sum^l_{i=0} \frac{1}{i! (l-i)! (l-1-i+b)!} \left( -\frac{m_c^2}{M^2} \right)^{l-i},
\label{F2}
\\
\mathcal{G}_1(n,a) &= \sum^{n-2}_{k=0} \frac{(-1)^k n! \Gamma(k+a) \Gamma(n-1-k)}{k! (n-k)! \Gamma(n-1+a)},
\label{G1}
\\
\mathcal{G}_2(n,a) &= \sum^{n-1}_{k=0} \frac{(-1)^k n! \Gamma(k+a) \Gamma(n-k)}{k! (n-k)! \Gamma(n+a)},
\label{G2}
\\
\mathcal{H}(n,a,b,c) &= \int^1_0 dx (2x-1)^n x^a (1-x)^b \exp \left[ -\frac{m_c^2}{M^2 (1-x)} \right]\nonumber\\
&= \frac{1}{(c-1)!} \frac{1}{(M^2)^c}\int^1_0 dx (2x-1)^n x^a (1-x)^b \exp \left[ -\frac{m_c^2}{M^2 (1-x)} \right].
\label{H}
\end{align}
And the Borel transformation formulas are,
\begin{align}
&
\hat{\cal B}_{M^2} \frac{1}{(-q^2 + m_c^2)^k} \ln \frac{-q^2 + m_c^2}{\mu^2} = \frac{1}{(k-1)!} \frac{1}{M^{2k}} e^{-m_c^2/M^2} \left[ \ln \frac{M^2}{\mu^2} + \psi(k) \right] \quad (k \geq 1),
\nonumber\\
&
\hat{\cal B}_{M^2} (-q^2 + m_c^2)^k \ln \frac{-q^2 + m_c^2}{\mu^2} = (-1)^{k+1} k! M^{2k} e^{-m_c^2/M^2} \quad (k \geq 0), \nonumber\\
&
\hat{\cal B}_{M^2} \dfrac{(-q^2)^l}{(-q^2+m_c^2)^{l+\tau}} = \left\{
\begin{array}{l l}
0, & \tau = 0, l = 0;\\[1.2ex]
\sum^{l-1}_{i=0} \dfrac{l!}{i! (l-i)! (l-i-1)!} \left( -\dfrac{m_c^2}{M^2} \right)^{l-i} e^{-m_c^2/M^2}, & \tau = 0, l > 0;\\
\sum^{l}_{i=0} \dfrac{l!}{i! (l-i)! (l+\tau-i-1)!} \left( -\dfrac{m_c^2}{M^2} \right)^{l-i} \dfrac{1}{M^{2\tau}} e^{-m_c^2/M^2}, & \tau > 0, l \geq 0.
\end{array}
\right.
\end{align}
\end{widetext}


\begin{thebibliography}{s2}

\bibitem{BABAR_Lees:2012xj}
  J.~P.~Lees {\it et al.} [BaBar Collaboration],
  Evidence for an excess of $\bar{B} \to D^{(\ast)} \tau^-\bar{\nu}_\tau$ decays,
  \href{https://doi.org/10.1103/PhysRevLett.109.101802}
  {Phys.\ Rev.\ Lett.\  {\bf 109}, 101802 (2012)}.

\bibitem{BABAR_Lees:2013uzd}
  J.~P.~Lees {\it et al.} [BaBar Collaboration],
  Measurement of an excess of $\bar{B} \to D^{(\ast)}\tau^- \bar{\nu}_\tau$ decays and implications for charged Higgs bosons,
  \href{https://doi.org/10.1103/PhysRevD.88.072012}
  {Phys.\ Rev.\ D {\bf 88}, 072012 (2013)}.

\bibitem{BELLE_Huschle:2015rga}
  M.~Huschle {\it et al.} [Belle Collaboration],
  Measurement of the branching ratio of $\bar{B} \to D^{(\ast)} \tau^- \bar{\nu}_\tau$ relative to $\bar{B} \to D^{(\ast)} \ell^- \bar{\nu}_\ell$ decays with hadronic tagging at Belle,
  \href{https://doi.org/10.1103/PhysRevD.92.072014}
  {Phys.\ Rev.\ D {\bf 92}, 072014 (2015)}.

\bibitem{HFAG_Amhis:2014hma}
  Y.~Amhis {\it et al.} [Heavy Flavor Averaging Group (HFAG)],
  Averages of $b$-hadron, $c$-hadron, and $\tau$-lepton properties as of summer 2014,
  \href{https://arxiv.org/abs/1412.7515}
  {  {arXiv:1412.7515}}.


\bibitem{LCSR_Zuo:2006dk}
  F.~Zuo, Z.~H.~Li and T.~Huang,
  Form factor for $B\to D l \tilde{\nu}$ in light-cone sum rules with chiral current correlator,
  \href{https://doi.org/10.1016/j.physletb.2006.07.039}
  {Phys.\ Lett.\ B {\bf 641}, 177 (2006)}.

\bibitem{Zuo:2006re}
  F.~Zuo and T.~Huang,
  $B_c(B)\to D\ell \tilde{\nu}$ form-factors in light-cone sum rules and the $D$ meson distribution amplitude,
  \href{https://doi.org/10.1088/0256-307X/24/1/017}
  {Chin.\ Phys.\ Lett.\ {\bf 24}, 61 (2007)}.

\bibitem{HQET_Fajfer:2012vx}
  S.~Fajfer, J.~F.~Kamenik and I.~Nisandzic,
  On the $B \to D^\ast \tau \bar{\nu}_{\tau}$ Sensitivity to New Physics,
  \href{https://doi.org/10.1103/PhysRevD.85.094025}
  {Phys.\ Rev.\ D {\bf 85}, 094025 (2012)}.

\bibitem{Fu:2013wqa}
  H.~B.~Fu, X.~G.~Wu, H.~Y.~Han, Y.~Ma and T.~Zhong,
  $|V_{cb}|$ from the semileptonic decay $B\to D \ell \bar{\nu}_{\ell}$ and the properties of the $D$ meson distribution amplitude,
  \href{https://doi.org/10.1016/j.nuclphysb.2014.04.021}
  {Nucl.\ Phys.\ B {\bf 884}, 172 (2014)}.

\bibitem{PQCD_Fan:2013qz}
  Y.~Y.~Fan, W.~F.~Wang, S.~Cheng and Z.~J.~Xiao,
  Semileptonic decays $B \to D^{(\ast)} \ell\nu$ in the perturbative QCD factorization approach,
  \href{https://doi.org/10.1007/s11434-013-0049-9}
  {Chin.\ Sci.\ Bull.\ {\bf 59}, 125 (2014)}.

\bibitem{PQCD_Fan:2015kna}
  Y.~Y.~Fan, Z.~J.~Xiao, R.~M.~Wang and B.~Z.~Li,
  The $B\to D^{(\ast)} \ell\nu_\ell$ decays in the pQCD approach with the Lattice QCD input,
  \href{https://doi.org/10.1007/s11434-015-0959-9}
  {Sci. Bull. {\bf 60}, 2009 (2015)}.

\bibitem{LQCD_Lattice:2015rga}
  J.~A.~Bailey {\it et al.} [MILC Collaboration],
  $B\to D\ell\nu$ form factors at nonzero recoil and $|V_{cb}|$ from 2+1-flavor lattice QCD,
  \href{https://doi.org/10.1103/PhysRevD.92.034506}
  {Phys.\ Rev.\ D {\bf 92}, 034506 (2015)}.

\bibitem{LQCD_Na:2015kha}
  H.~Na {\it et al.} [HPQCD Collaboration],
  $B \to D \ell \nu$ form factors at nonzero recoil and extraction of $|V_{cb}|$,
  \href{https://doi.org/10.1103/PhysRevD.93.119906}
  {Phys.\ Rev.\ D {\bf 92}, 054510 (2015)}.

\bibitem{Wang:2017jow}
  Y.~M.~Wang, Y.~B.~Wei, Y.~L.~Shen and C.~D.~L\"u,
  Perturbative corrections to $B\to D$ form factors in QCD,
  \href{https://doi.org/10.1007/JHEP06(2017)062}
  {JHEP {\bf 1706}, 062 (2017)}.

\bibitem{Zhong:2018exo}
  T.~Zhong, Y.~Zhang, X.~G.~Wu, H.~B.~Fu and T.~Huang,
  The ratio $\mathcal {R}(D)$ and the $D$-meson distribution amplitude,
  \href{https://doi.org/10.1140/epjc/s10052-018-6387-7}
  {Eur.\ Phys.\ J.\ C {\bf 78}, 937 (2018)}.

\bibitem{Aaij:2020hsi}
  R.~Aaij \textit{et al.} [LHCb Collaboration],
  Measurement of $|V_{cb}|$ with $B_s^0 \to D_s^{(*)-} \mu^+ \nu_{\mu}$ decays,
  \href{https://doi.org/10.1103/PhysRevD.101.072004}
  {Phys. Rev. D \textbf{101} (2020) 072004}.

\bibitem{Caprini:1997mu}
  I.~Caprini, L.~Lellouch and M.~Neubert,
  Dispersive bounds on the shape of $\bar{B}\to D^{(\ast)}\ell\bar{\nu}$ form-factors,
  \href{https://doi.org/10.1016/S0550-3213(98)00350-2}
  {Nucl. Phys. B \textbf{530} (1998) 153}.

\bibitem{Boyd:1994tt}
  C.~G.~Boyd, B.~Grinstein and R.~F.~Lebed,
  Constraints on form-factors for exclusive semileptonic heavy to light meson decays,
  \href{https://doi.org/10.1103/PhysRevLett.74.4603}
  {Phys.\ Rev.\ Lett.\  {\bf 74}, 4603 (1995)}.

\bibitem{Boyd:1995sq}
  C.~G.~Boyd, B.~Grinstein and R.~F.~Lebed,
  Model independent determinations of $\bar B \to D \ell\bar\nu, D^* \ell\bar\nu$ form-factors,
  \href{https://doi.org/10.1016/0550-3213(95)00653-2}
  {Nucl.\ Phys.\ B {\bf 461}, 493 (1996)}.

\bibitem{Boyd:1997kz}
  C.~G.~Boyd, B.~Grinstein and R.~F.~Lebed,
  Precision corrections to dispersive bounds on form-factors,
  \href{https://doi.org/10.1103/PhysRevD.56.6895}
  {Phys.\ Rev.\ D {\bf 56}, 6895 (1997)}.

\bibitem{Faustov:2012mt}
  R.~N.~Faustov and V.~O.~Galkin,
  Weak decays of $B_s$ mesons to $D_s$ mesons in the relativistic quark model,
  \href{https://doi.org/10.1103/PhysRevD.87.034033}
  {Phys.\ Rev.\ D {\bf 87}, 034033 (2013)}.

\bibitem{Kramer:1992xr}
  G.~Kramer and W.~F.~Palmer,
  Decay of $B_{s}$ mesons into vector mesons,
  \href{https://doi.org/10.1103/PhysRevD.46.3197}
  {Phys.\ Rev.\ D {\bf 46}, 3197 (1992)}.

\bibitem{Chen:2011ut}
  X.~J.~Chen, H.~F.~Fu, C.~S.~Kim and G.~L.~Wang,
  Estimating Form Factors of $B_s\rightarrow D_s^{(\ast)}$ and their Applications to Semi-leptonic and Non-leptonic Decays,
\href{https://doi.org/10.1088/0954-3899/39/4/045002}
  {J.\ Phys.\ G {\bf 39}, 045002 (2012)}.

\bibitem{Blasi:1993fi}
  P.~Blasi, P.~Colangelo, G.~Nardulli and N.~Paver,
  Phenomenology of $B_s$ decays,
  \href{https://doi.org/10.1103/PhysRevD.49.238}
  {Phys.\ Rev.\ D {\bf 49}, 238 (1994)}.

\bibitem{Li:2009wq}
  R.~H.~Li, C.~D.~L\"u and Y.~M.~Wang,
  Exclusive $B_s$ decays to the charmed mesons $D^+_{s}$(1968,2317) in the standard model,
  \href{https://doi.org/10.1103/PhysRevD.80.014005}
  {Phys.\ Rev.\ D {\bf 80}, 014005 (2009)}.

\bibitem{Atoui:2013zza}
  M.~Atoui, V.~Mor\'{e}nas, D.~Be\v{c}irevic and F.~Sanfilippo,
  $B_{s} \to D_{s}\ell\nu_\ell$  near zero recoil in and beyond the Standard Model,
  \href{https://doi.org/10.1140/epjc/s10052-014-2861-z}
  {Eur.\ Phys.\ J. \ C {\bf 74}, 2861 (2014)}.

\bibitem{Monahan:2017uby}
  C.~J.~Monahan, H.~Na, C.~M.~Bouchard, G.~P.~Lepage and J.~Shigemitsu,
  $B_s \to D_s \ell \nu$ Form Factors and the Fragmentation Fraction Ratio $f_s/f_d$,
  \href{https://doi.org/10.1103/PhysRevD.95.114506}
  {Phys.\ Rev.\ D {\bf95}, 114506 (2017)}.

\bibitem{McLean:2019qcx}
  E.~McLean, C.~T.~H.~Davies, J.~Koponen and A.~T.~Lytle,
  $B_s\to D_s \ell\nu$ Form Factors for the full $q^2$ range from Lattice QCD with non-perturbatively normalized currents,
  \href{https://doi.org/10.1103/PhysRevD.101.074513}
  {Phys.\ Rev.\ D {\bf 101}, 074513 (2020)}.

\bibitem{Huang:2004hw}
  T.~Huang and X.~G.~Wu,
  Consistent calculation of the B to pi transition form-factor in the whole physical region,
  \href{https://doi.org/10.1103/PhysRevD.71.034018}
  {Phys.\ Rev.\ D {\bf 71}, 034018 (2005)}.

\bibitem{Huang:1998gp}
  T.~Huang and Z.~H.~Li,
  ``$B \to K^*$ gamma in the light cone QCD sum rule,''
  \href{https://doi.org/10.1103/PhysRevD.57.1993}
  {Phys.\ Rev.\ D {\bf 57}, 1993 (1998)}.

\bibitem{Huang:2001xb}
  T.~Huang, Z.~H.~Li and X.~Y.~Wu,
  ``Improved approach to the heavy to light form-factors in the light cone QCD sum rules,''
  \href{https://doi.org/10.1103/PhysRevD.63.094001}
  {Phys.\ Rev.\ D {\bf 63}, 094001 (2001)}.

\bibitem{Wan:2002hz}
  Z.~G.~Wang, M.~Z.~Zhou and T.~Huang,
  ``$B \pi$ weak form-factor with chiral current in the light cone sum rules,''
  \href{https://doi.org/10.1103/PhysRevD.67.094006}
  {Phys.\ Rev.\ D {\bf 67}, 094006 (2003)}.

\bibitem{Zuo:2006dk}
  F.~Zuo, Z.~H.~Li and T.~Huang,
  ``Form Factor for $B \to D l \nu$ in Light-Cone Sum Rules With Chiral Current Correlator,''
  \href{https://doi.org/10.1016/j.physletb.2006.07.039}
  {Phys.\ Lett.\ B {\bf 641}, 177 (2006)}.

\bibitem{Wu:2007vi}
  X.~G.~Wu, T.~Huang and Z.~Y.~Fang,
  ``SU(f)(3)-symmetry breaking effects of the $B \to K$ transition form-factor in the QCD light-cone sum rules,''
  \href{https://doi.org/10.1103/PhysRevD.77.074001}
  {Phys.\ Rev.\ D {\bf 77}, 074001 (2008)}.

\bibitem{Wu:2009kq}
  X.~G.~Wu and T.~Huang,
  ``Radiative Corrections on the $B \to P$ Form Factors with Chiral Current in the Light-Cone Sum Rules,''
  \href{https://doi.org/10.1103/PhysRevD.79.034013}
  {Phys.\ Rev.\ D {\bf 79}, 034013 (2009)}.

\bibitem{Dhiman:2019ddr}
  N.~Dhiman, H.~Dahiya, C.~R.~Ji and H.~M.~Choi,
  Twist-2 pseudoscalar and vector meson distribution amplitudes in light-front quark model with exponential-type confining potential,
  \href{https://doi.org/10.1103/PhysRevD.100.014026}
  {Phys. Rev. D \textbf{100} (2019), 014026}.

\bibitem{BHL1}
  S. J. Brodsky, T. Huang, and G. P. Lepage, in \textit{Particles and Fields-2}, Proceedings of the Banff Summer Institute, Banff; Alberta, 1981, edited by A. Z. Capri and A. N. Kamal (Plenum, New York, 1983), p. 143;

\bibitem{BHL2}
  G. P. Lepage, S. J. Brodsky, T. Huang, and P. B.Mackenize, in \textit{Particles and Fields-2}, Proceedings of the Banff Summer Institute, Banff; Alberta, 1981, edited by A. Z. Capri and A. N. Kamal (Plenum, New York, 1983), p. 83;

\bibitem{BHL3}
  T. Huang, in \textit{Proceedings ofXXth International Conference on High Energy Physics}, Madison, Wisconsin, 1980, edited by L. Durand and L. G Pondrom, AIP Conf. Proc. No. 69 (AIP, New York, 1981), p. 1000.

\bibitem{BHL_Zhong:2014jla}
  T.~Zhong, X.~G.~Wu, Z.~G.~Wang, T.~Huang, H.~B.~Fu and H.~Y.~Han,
  Revisiting the pion leading-twist distribution amplitude within the QCD background field theory,
  \href{https://doi.org/10.1103/PhysRevD.90.016004}
  {Phys.\ Rev.\ D {\bf 90}, 016004 (2014)}.

\bibitem{BHL_Zhong:2014fma}
  T.~Zhong, X.~G.~Wu and T.~Huang,
  Heavy pseudoscalar leading-twist distribution amplitudes within QCD theory in background fields,
  \href{https://doi.org/10.1140/epjc/s10052-015-3271-6}
  {Eur.\ Phys.\ J.\ C {\bf 75}, 45 (2015)}.

\bibitem{BHL_Zhong:2016kuv}
  T.~Zhong, X.~G.~Wu, T.~Huang and H.~B.~Fu,
  Heavy pseudoscalar twist-3 distribution amplitudes within QCD theory in background fields,
  \href{https://doi.org/10.1140/epjc/s10052-016-4350-z}
  {Eur.\ Phys.\ J.\ C {\bf 76}, 509 (2016)}.

\bibitem{BHL_Zhang:2017rwz}
  Y.~Zhang, T.~Zhong, X.~G.~Wu, K.~Li, H.~B.~Fu and T.~Huang,
  Uncertainties of the $B\to D$ transition form factor from the $D$-meson leading-twist distribution amplitude,
  \href{https://doi.org/10.1140/epjc/s10052-018-5551-4}
  {Eur.\ Phys.\ J.\ C {\bf 78}, 76 (2018)}.

\bibitem{BHL_Zhong:2018exo}
  T.~Zhong, Y.~Zhang, X.~G.~Wu, H.~B.~Fu and T.~Huang,
  The ratio $\mathcal {R}(D)$ and the $D$-meson distribution amplitude,
  \href{https://doi.org/10.1140/epjc/s10052-018-6387-7}
  {Eur.\ Phys.\ J.\ C {\bf 78}, 937 (2018)}.

\bibitem{Fu:2014cna}
  H.~B.~Fu, X.~G.~Wu, H.~Y.~Han, Y.~Ma and H.~Y.~Bi,
  The $\rho$-meson longitudinal leading-twist distribution amplitude,
  \href{https://doi.org/10.1016/j.physletb.2014.09.055}
  {Phys. Lett. B \textbf{738}, 228 (2014)}.

\bibitem{Fu:2014pba}
  H.~B.~Fu, X.~G.~Wu, H.~Y.~Han and Y.~Ma,
  $B \to \rho$ transition form factors and the $\rho$-meson transverse leading-twist distribution amplitude,
  \href{https://doi.org/10.1088/0954-3899/42/5/055002}
  {J. Phys. G \textbf{42}, 055002 (2015)}.

\bibitem{Zeng:2021hwt}
  L.~Zeng, H.~B.~Fu, D.~D.~Hu, L.~L.~Chen, W.~Cheng and X.~G.~Wu,
  Revisiting the production of $J/\psi+\eta_c$ via the $e^+e^-$ annihilation within the QCD light-cone sum rules,
  \href{https://arxiv.org/abs/2102.01842}
  {  {arXiv:2102.01842}}.

\bibitem{SVZ_Shifman:1978bx}
  M.~A.~Shifman, A.~I.~Vainshtein and V.~I.~Zakharov,
  QCD and Resonance Physics. Theoretical Foundations,
  \href{https://doi.org/10.1016/0550-3213(79)90022-1}
  {Nucl.\ Phys.\ B {\bf 147}, 385 (1979)}.

\bibitem{BFT_Huang:1989gv}
  T.~Huang and Z.~Huang,
  Quantum Chromodynamics in Background Fields,
  \href{https://doi.org/10.1103/PhysRevD.39.1213}
  {Phys.\ Rev.\ D {\bf 39}, 1213 (1989)}.

\bibitem{Duplancic:2008ix}
  G.~Duplancic, A.~Khodjamirian, T.~Mannel, B.~Melic and N.~Offen,
  Light-cone sum rules for $B \to \pi$ form factors revisited,
  \href{https://doi.org/10.1088/1126-6708/2008/04/014}
  {JHEP \textbf{04} (2008), 014}. The $s$-quark mass is small in comparison to the $b$-quark mass, so we neglect the $s$-quark mass effects in deriving the $B_{s}\to D_{s}$ TFF. Then the imaginary part of the NLO amplitude $T_1$ is the same as that of $B\to\pi$ TFF.

\bibitem{Zhong:2021}
  T. Zhong, Z. H. Zhu, H. B. Fu, X. G. Wu and T. Huang,
  An improved light-cone harmonic oscillator model for the pionic leading-twist distribution amplitudes.
  \href{https://arxiv.org/abs/2102.03989}
  {  {arXiv:2102.03989}}.

\bibitem{Hu:2021zmy}
  D.~D.~Hu, H.~B.~Fu, T.~Zhong, L.~Zeng, W.~Cheng and X.~G.~Wu,
  $\eta$-meson leading-twist distribution amplitude within QCD sum rule approach and its application to the semi-leptonic decay $ D_s^+ \to\eta{\ell}^{+} \nu_{\ell}$,
  \href{https://arxiv.org/abs/2102.05293}
  {  {arXiv:2102.05293}}.

\bibitem{Huang:1994dy}
  T.~Huang, B.~Q.~Ma and Q.~X.~Shen,
  Analysis of the pion wave function in light cone formalism,
  \href{https://doi.org/10.1103/PhysRevD.49.1490}
  {Phys.\ Rev.\ D {\bf 49}, 1490 (1994)}.

\bibitem{Lepage:1980fj}
  G.~P.~Lepage and S.~J.~Brodsky,
  Exclusive Processes in Perturbative Quantum Chromodynamics,
  \href{https://doi.org/10.1103/PhysRevD.22.2157}
  {Phys.\ Rev.\ D {\bf 22}, 2157 (1980)}.

\bibitem{MOLELIV_Guo:1991eb}
  X.~H.~Guo and T.~Huang,
  Hadronic wave functions in $D$ and $B$ decays,
  \href{https://doi.org/10.1103/PhysRevD.43.2931}
  {Phys.\ Rev.\ D {\bf 43}, 2931 (1991)}.

\bibitem{Zyla:2020zbs}
  P.A.~Zyla \textit{et al.} [Particle Data Group],
  Review of Particle Physics,
  \href{https://doi.org/10.1093/ptep/ptaa104}
  {PTEP \textbf{2020} (2020), 083C01}.

\bibitem{SRREV_Colangelo:2000dp}
  P.~Colangelo and A.~Khodjamirian,
  QCD sum rules, a modern perspective,
  \href{https://doi.org/10.1142/9789812810458\_0033}
  {At the frontier of particle physics, vol. 3* 1495-1576}.

\bibitem{Narison:2014wqa}
  S.~Narison,
  Mini-review on QCD spectral sum rules,
  \href{https://doi.org/10.1016/j.nuclphysbps.2015.01.041}
  {Nucl.\ Part.\ Phys.\ Proc.\ {\bf 258-259}, 189 (2015)}.

\bibitem{RGE_Yang:1993bp}
  K.~C.~Yang, W.~Y.~P.~Hwang, E.~M.~Henley and L.~S.~Kisslinger,
  QCD sum rules and neutron proton mass difference,
  \href{https://doi.org/10.1103/PhysRevD.47.3001}
  {Phys.\ Rev.\ D {\bf 47}, 3001 (1993)}

\bibitem{RGE_Hwang:1994vp}
  W.~Y.~P.~Hwang and K.~C.~Yang,
  QCD sum rules: $\Delta-N$ and $\Sigma_0-\Lambda$ mass splittings,
  \href{https://doi.org/10.1103/PhysRevD.49.460}
  {Phys.\ Rev.\ D {\bf 49}, 460 (1994)}

\bibitem{Wang:2008xt}
  W.~Wang, Y.~L.~Shen and C.~D.~L\"u,
  Covariant Light-Front Approach for $B_{(c)}$ transition form factors,
  \href{https://doi.org/10.1103/PhysRevD.79.054012}
  {Phys. Rev. D \textbf{79}, 054012 (2009)}.

\bibitem{Aubert:2008yv}
  B.~Aubert \textit{et al.} [BaBar Collaboration],
  Measurements of the Semileptonic Decays $\bar B \to D \ell \bar \nu$ and $\bar B \to D^* \ell \bar \nu$ Using a Global Fit to $D X \ell \bar \nu$ Final States,
  \href{https://doi.org/10.1103/PhysRevD.79.012002}
  {Phys. Rev. D \textbf{79} (2009), 012002}.

\bibitem{Aubert:2009ac}
  B.~Aubert \textit{et al.} [BaBar Collaboration],
  Measurement of $|V_{cb}|$ and the form-factor slope in $\bar B \to D \ell \nu_\ell$ decays in events tagged by a fully reconstructed $B$ meson,
  \href{https://doi.org/10.1103/PhysRevLett.104.011802}
  {Phys. Rev. Lett. \textbf{104} (2010), 011802}.

\bibitem{Dey:2019bgc}
  J.~P.~Lees \textit{et al.} [BaBar Collaboration],
  Extraction of form Factors from a Four-Dimensional Angular Analysis of $\overline{B} \rightarrow D^\ast \ell^- \overline{\nu}_\ell$,
  \href{https://doi.org/10.1103/PhysRevLett.123.091801}
  {Phys. Rev. Lett. \textbf{123} (2019), 091801}.

\bibitem{Waheed:2018djm}
  E.~Waheed \textit{et al.} [Belle Collaboration],
  Measurement of the CKM matrix element $|V_{cb}|$ from $B^0\to D^{*-}\ell^ {+} \nu_\ell$ at Belle,
  \href{https://doi.org/10.1103/PhysRevD.100.052007}
  {Phys. Rev. D \textbf{100} (2019) no.5, 052007}.

\bibitem{Harrison:2017fmw}
  J.~Harrison \textit{et al.} [HPQCD Collaboration],
  Lattice QCD calculation of the ${{B}_{(s)}\to D_{(s)}^{*}\ell{\nu}}$ form factors at zero recoil and implications for ${|V_{cb}|}$,
  \href{https://doi.org/10.1103/PhysRevD.97.054502}
  {Phys. Rev. D \textbf{97} (2018), 054502}.

\end{thebibliography}
\end{document}